\documentclass{cernyrep}
\usepackage{texnames}
\usepackage[T1]{fontenc}
\usepackage{varwidth}
\usepackage{xcolor}
\usepackage[bookmarks, colorlinks=true, linktoc=page, pdftex, linkcolor=black, citecolor=black, urlcolor=blue]{hyperref}
\usepackage[ocgcolorlinks]{ocgx2}

\usepackage{listings}

\usepackage{fancyhdr}
\fancyhfoffset{4 mm}
\fancypagestyle{ARTTITLE}{%
\fancyhf{} 
\lhead{\small{Proceedings of the 2018 CERN--Accelerator--School course on\\
\it{Numerical Methods for Analysis, Design and Modelling of Particle Accelerators}, Thessaloniki, (Greece)}} 
\lfoot{Available online at \url{https://cas.web.cern.ch/previous-schools}}
\rfoot{\thepage\hspace*{3mm}}
 
}
\renewcommand{\headheight}{22.38pt}

\begin{document}
\title{Computing techniques}

\author{X. Buffat}

\institute{CERN, Geneva, Switzerland}

\maketitle 
\thispagestyle{ARTTITLE}

\begin{abstract}
This lecture aims at providing a user's perspective on the main concepts used nowadays for the implementation of numerical algorithm on common computing architecture. In particular, the concepts and applications of Central Processing Units (CPUs), vectorisation, multithreading, hyperthreading and Graphical Processing Units (GPUs), as well as computer clusters and grid computing will be discussed. Few examples of source codes illustrating the usage of these technologies are provided.
\end{abstract}

\section{Introduction}
Nowadays several problems can be addressed only via numerical algorithms. The study of particle accelerators is clearly among the many fields relying heavily on highly performing computers to execute various types of numerical algorithms for design, measurement, correction and operation of the machines and their individual components as described in the various contributions to these proceedings. Conceptually, all these algorithms consist of a series of operations to be executed on a given dataset. In practice, the number of these operation as well as the size of the dataset is limited by the available technologies. Here we aim at describing the main concepts that are needed to understand the current technologies commonly used for the implementation of numerical algorithms, with emphasis on the accelerator world. This field is rather wide, this lecture is mainly designed for beginners in computing wishing to broaden their knowledge in view of using more efficiently their personal computers or to investigate which high performance computing technologies would be most suited for a given application.

In Sec.~\ref{sec-CPU}, we describe the main components of the Central Processing Unit (CPU) of the most common computer, i.e. based on transistors, and the basics of the corresponding nomenclature. The possibility to combine the power of multiple CPUs to achieve a common task is discussed in Sec.~\ref{sec-multCPU}. Section~\ref{sec-amdhal} describes the improvements of the performance that can be expected using parallelism. We'll then consider an evolution of the CPU available commercially, aiming at massively parallel computations, i.e. the Graphical Processing Unit (GPU) (Sec.~\ref{sec-GPU}). Eventually, the combination of multiple computers to achieve large tasks will be discussed in Secs.~\ref{sec-cluster} and~\ref{sec-grid}, with the concept of computer clusters, volunteer computing and computing grids.
\section{The central processing unit} \label{sec-CPU}
\begin{figure}
 \begin{center}
  \includegraphics[width=0.8\linewidth]{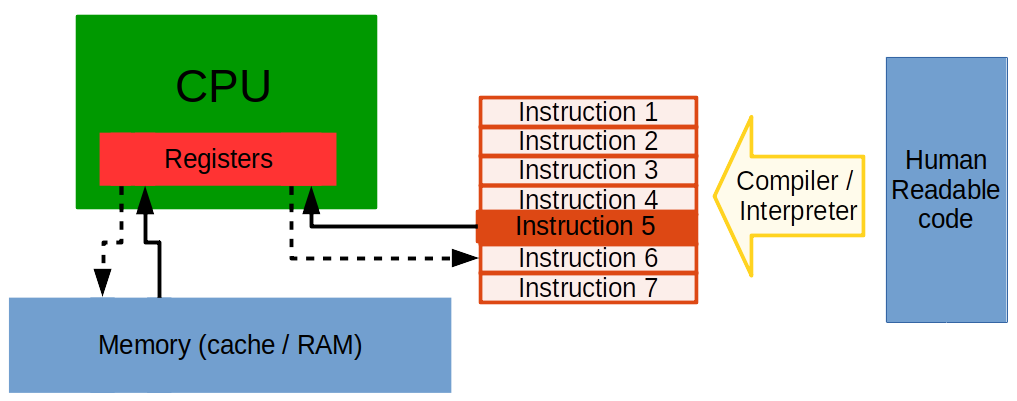}
 \end{center}
\caption{Sketch of the main steps of the execution of a CPU cycle. A given list of instruction is generated by a compiler or an interpreter based on the source code. At each step, a given instruction and the corresponding data are loaded in the registers (solid lines), the operation is executed by the CPU, and the result is copied back into the memory or the next instructions are modified depending on the type of instruction executed (dashed lines).}
\label{fig-CPUIllustration}
\end{figure}
From the user's point of view, a CPU can be viewed as a device capable of performing a given set of operations on a given data set. The CPU is operated in cycles during which one of the instruction from the set is executed, as illustrated in Fig.~\ref{fig-CPUIllustration}. The main characteristic of the CPU is then the number of instructions that can be performed per unit time, usually referred to as the clock speed, which nowadays is limited to a few~GHz.

Most CPUs implement mathematical operations ranging from the simple arithmetic or comparison operators to more complex functions such as trigonometric or exponential functions. The capabilities of a given CPU are described by the instruction sets that they implement. Most of the CPUs in nowadays personal computers implement the so-called \textit{x86-64} instruction set featuring the operations required for most applications in accelerator physics, more generally for scientific computing. On the other hand, the CPUs in smart phones or tablets implement other sets of instructions, making them less performant for scientific computing.

Each instruction applies to a given data type. This data has to be stored in a dedicated memory on the CPU, called the register. During each of its cycles, the CPU will perform a given instruction to the data stored in the register. We note that the instruction itself is also encoded in another dedicated register on the CPU. For example, the operation could be the replacement of two double precision numbers each represented by 64 bits by their sum, also represented by 64 bits. To execute such an operation, the register has to be of at least 128~bits. The size of the register is therefore an important characteristic of the CPU, it will be further discussed in the next section.

While the CPU instructions can be addressed directly even in high level programming languages, via so-called intrinsics~\cite{intrinsics}, their direct usage is rather uncommon. The building of the list of instructions to be executed by the CPU is usually done by a compiler or an interpreter, based on a human readable source code. The role of the compiler or the interpreter is therefore crucial in the performance of a given computer code.

For completeness we note that, on top of the mathematical operations, the capabilities of the CPU include the transfer of the data from the computer's memory to the register and back, as well as the modifications of the following instructions that will be assigned to it in the next cycles (e.g. an \textit{if} statement leading to the execution of one or another part of the code, i.e. another list of instructions).

\subsection{Vectorisation} \label{sec-vector}
Above we took the example of the addition of two numbers represented by 64 bits each, thus requiring at least a 128 bits register. For many implications, the same operations has to be performed on several elements of an array, in that case an improvement of the speed of the execution by a factor $N_v$ can be obtained by loading $N_v$ pairs of numbers in the register and perform the addition of all the pairs in one cycle. To allow for such an operation, the size of the register has to be at least $N_v\cdot$128 bits. This form of parallelism is known as vectorisation.

The vectorisation capabilities are evolving significantly nowadays, e.g. the Intel Xeon Phi features a 512~bits register, thus allowing for an eight-fold vectorisation for double precision operations. The vectorisation capability of a given CPU can be inferred from the standard instruction sets that it implements: AVX and AVX-2 correspond to 256~bits registers and AVX-512 to 512 bits~\cite{intrinsics}.

The code developer can access vectorised instructions via the corresponding intrinsics, however the most common usage is to rely on the compiler or the interpreter to perform the right choice of instructions. For most compilers the vectorisation is not enabled by default. For \textit{gcc}, the flag \textit{-ftree-vectorize} (implied by the optimisation flag \textit{-O3}) enables vectorisation~\cite{gcc}. However the capability of the compiler to recognise vectorisable loops is limited in particular for complicated ones. Thus, while this parallelisation does in principle not require any modification of the source code, it favours the implementation of simple loops for an efficient execution of the heavy tasks.
\section{Multiple CPUs}\label{sec-multCPU}
The vectorisation goes under the classification of Single Instruction, Multiple Data (SIMD) parallelisation, which are mostly suited for homogeneous tasks to be performed on large datasets. In many cases however the tasks are heterogeneous and such a parallelism becomes insufficient. In such cases, it is more suited to perform Multiple Instruction on Multiple Data (MIMD). On common computers, this is achieved by affecting multiple CPUs to a given task in an organised way, using the concept of threads.
\subsection{Multithreading}
So-called serial codes considered in the previous section feature a single sequence of instructions to be performed by a single CPU. In order to allow for multiple CPUs to perform different tasks simultaneously one requires separate lists of instructions for each of the CPUs, so-called threads. A certain level of synchronisation and communication between the threads is ensured via an access to a shared memory which is discussed in more detail in the next section. Conceptually, the number of threads is independent of the number of CPUs on which the program is executed. Multithreaded source codes can in principle be written without any knowledge of the architecture. If it occurs that there are more threads than CPUs, the threads will share the available resources, having consequences on the memory, and consequently on the performance (Sec.~\ref{sec-numa}). Since the operating systems is responsible for the allocation of the resources to the different threads, his role can be crucial in the performance of a multithreaded code. In some cases, a dedicated setting up of the operating system can be necessary, an example is discussed in the next sections.

An example of \textit{C} source code featuring multithreading using POSIX threads~\cite{POSIX} is shown in Sec.~\ref{sec-pthreads}. This simple example does not demonstrate the flexibility of POSIX threads, yet it features the two most basic elements, i.e. the creation and synchronisation of the threads. At their creation (lines 59 to 62), the four additional threads will execute independently, possibly on different CPUs if available, until their synchronisation (lines 65 to 68). This parallelisation is highly efficient since there is no copy of the memory, i.e. all of the threads act on the same memory. This efficiency comes with the responsibility to ensure that the threads do not interfere with each other, through the common memory. This can be achieved even for complex tasks through several concepts such as critical sections, reductions, atomic operations, locks or barriers. These concepts are detailed for example in~\cite{multithreading}.

While the POSIX threads allows for the implementation of various complex tasks, their usage requires the usage of a given syntax in the source code which differs significantly from the equivalent serial code. The Application Programming Interface (API) OpenMP~\cite{openMP} is particularly handy for the parallelisation of loops using multithreading without effort on the source code. An example of \textit{C} source code parallelised with OpenMP is shown in Sec.~\ref{sec-openMP}. This code is in fact a serial code, for which a single statement has been added (lines 33-35). This so-called \textit{pragma} instructs the compiler to parallelise the following loop using multiple threads. This method effortlessly generates a significant speed-up for most applications even on a common desktop computer featuring few CPUs. For more demanding applications, high performance computers featuring few tens of CPUs with a shared memory are commercially available nowadays.

In the accelerator world, this type of parallelism is not only used for scientific computing but is also particularly suited for controls software. The independent threads can be dedicated to the control of several individual systems interacting together through a single application.
\subsection{Cache memories}\label{sec-numa}
\begin{figure}
 \begin{center}
  \includegraphics[width=0.9\linewidth]{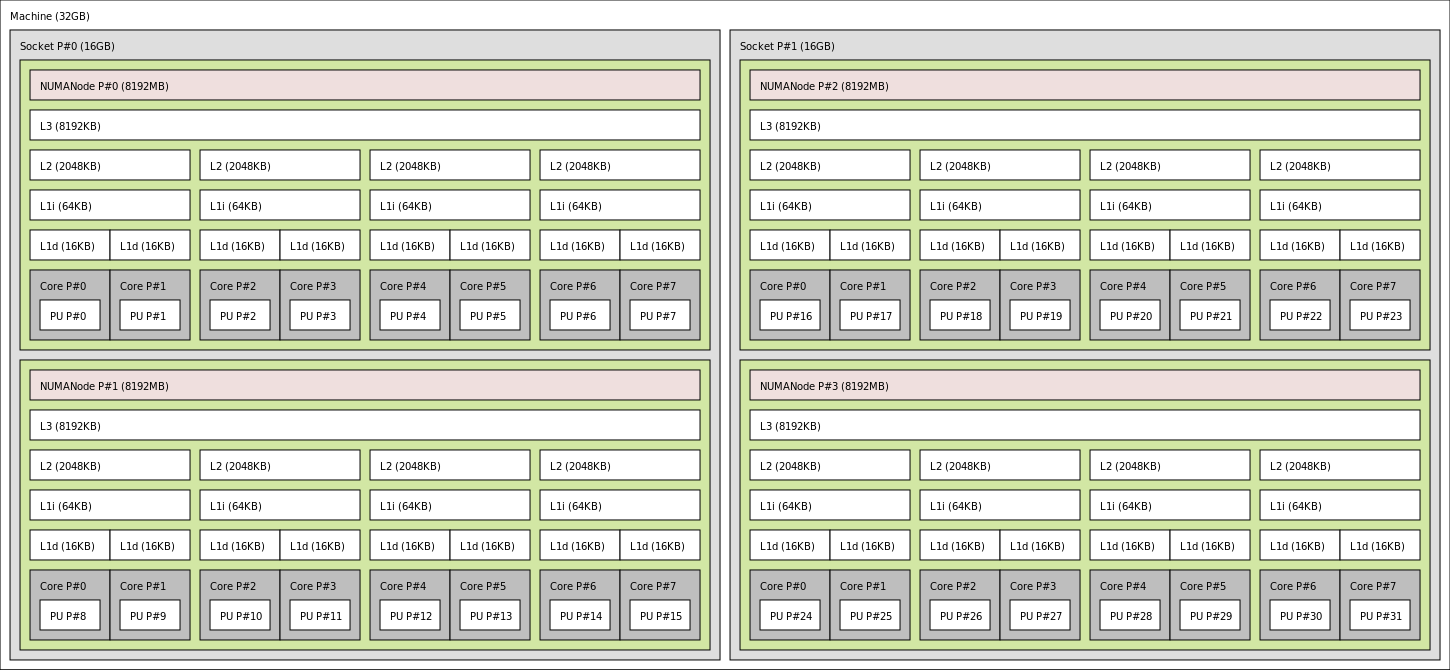}
 \end{center}
\caption{Topology of the cache memories on a machine with two sockets each featuring two NUMA zones with 8 cores and four levels of cache. Courtesy~\cite{wiki-numa}.}
\label{fig-NUMA}
\end{figure}
When increasing the number of CPUs, the performance of multithreaded codes can become limited by the transfer of data between CPUs, as it is technologically challenging to implement large memories close to the CPU, i.e. with a fast interconnection. In order to maximise the efficiency of the data transfer to the CPUs' registers, cache memories are implemented close to the CPUs, thus allowing for less frequent transfers to the main computer memory which are significantly slower. In fact, the caches are sub-divided in various levels, from small and close ones to larger and further ones. The example of such a hierarchical implementation is shown in Fig.~\ref{fig-NUMA}.

While the optimisation of a code for a given architecture is beyond the scope of this lecture, it is good to keep in mind that the strategy of the allocation of the memory has an impact on the performance. A simple example: for applications in which each thread acts mainly on a given fraction of the dataset, such as the application illustrated in Sec.~\ref{sec-sources}, it is advantageous for the data to be allocated close to the CPU executing the thread. Consequently, the performance is expected to be optimal if the execution of the threads are constrained to a given CPU. This is not the behaviour by default for most operating systems, but can be setup when needed.
\subsection{Simultaneous multithreading (Hyperthreading)}
The time required for the transfer of the data between the CPU registers and the memory (i.e. either the caches or the main memory) can be significantly longer than a CPU cycle. During this time, the CPU is stalled for potentially several cycles. In order to profit from this downtime, modern CPUs feature additional registers allowing them to hold the data required for the execution of two threads, yet a single instruction can be executed at once for one of them. The so-called simultaneous multithreading is mostly known under the name of Intel's implementation: Hyperthreading. While each CPU equipped with this technology appears as a separate core from the operating system point of view, the performance is usually between the equivalent of 1 to 2 separate CPUs. For example, the code shown in Sec.~\ref{sec-openMP} executed in 277~s serially and in 15~s with OpenMP enabled with 24 threads executing on 12 2.3~GHz Intel Xeon CPUs with hyperthreading enabled, thus generating a speed-up of $\approx$18.
\section{Amdhal's law}\label{sec-amdhal}

\begin{figure}
 \begin{center}
  \includegraphics[width=0.7\linewidth]{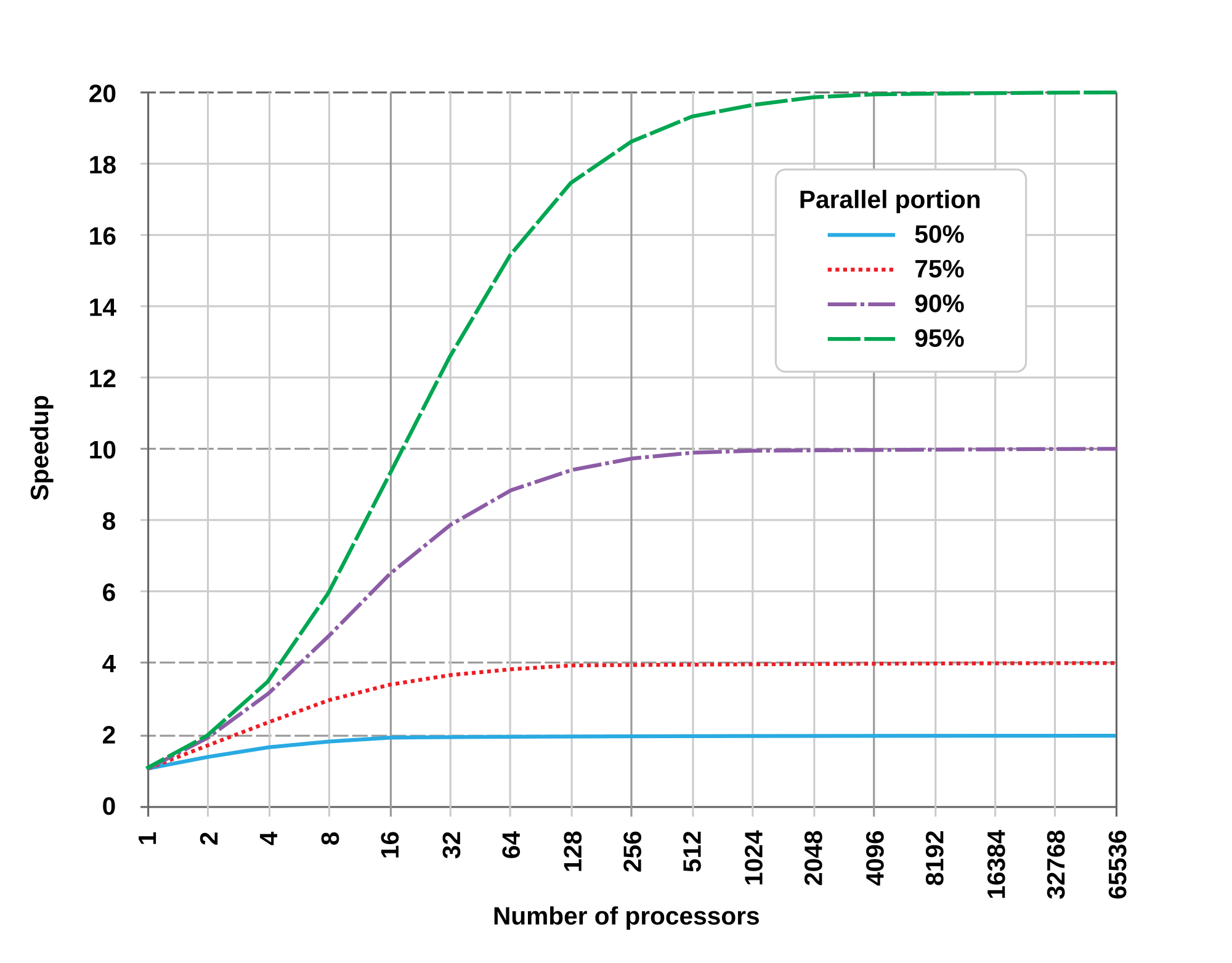}
 \end{center}
\caption{Illustration of the saturation of the speed up that can be achieved by parallelisation due to the fraction of non-parallelisable code, namely Amdhal's law. Courtesy~\cite{wiki-amdhal}.}
\label{fig-amdhal}
\end{figure}
Leaving aside memory limitations such as those discussed in the previous section, the execution of a parallelised code remains limited by its fraction of non-parallelisation code. If we denote this fraction of the execution $\alpha$, only the execution of the remaining fraction $(1-\alpha)$ is reduced to $(1-\alpha)/N_c$ by a parallelisation with $N_c$ cores. Thus we can write the total speed up~\cite{amdhal}:
\begin{equation}
 S = \frac{1}{\alpha+\frac{1-\alpha}{N_c}}.
\end{equation}
This so-called Amdhal's law quantifies the saturation of the performance of parallelised codes due to the non-parallelisable part (Fig.~\ref{fig-amdhal}). It is useful on one side to estimate the resources worth investing into the solution of a given problem as well as for diagnosing a given implementation of a parallelised code.

\section{The graphical processing unit}  \label{sec-GPU}
\begin{figure}
 \begin{center}
  \includegraphics[width=0.8\linewidth]{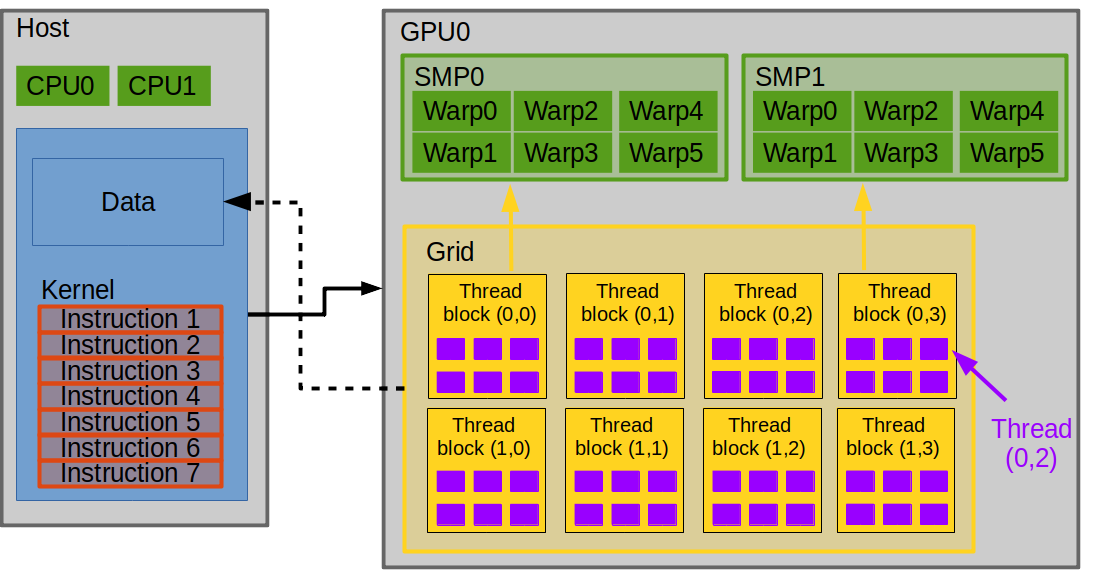}
 \end{center}
\caption{Illustration of the operation of a GPU. The execution of a kernel on a GPU card consists first of the loading of the instructions as well as the corresponding data to the card memory (solid black line). Then the threads are loaded in the SMPs caches and executed (here illustrated by one yellow arrow per SMP, each executing the threads in a given thread block). Finally the data is transferred back from the card memory to the host memory (dashed black line).}
\label{fig-GPU}
\end{figure}
The vectorisation (Sec.~\ref{sec-vector}) is particularly suited for the processing of images and videos, where the processing of large amounts of quasi-independent data is required, leading to the development of a dedicated technology, the GPU. This technology combines large vectorisation capabilities with concepts similar to multithreading and hyperthreading to achieve a number of operations per second that currently exceeds those of CPUs by one to two orders of magnitude. This technology is therefore suited for other applications featuring a similar level independence in the data as in image processing.

While some technologies enable the usage of GPUs without effort from the developers point of view (e.g. see Sec.~\ref{sec-openACC} or via the usage of libraries implementing GPU acceleration), an efficient usage of the GPU requires some understanding of its operation and often the usage of dedicated languages.

As the architecture of a GPU differs significantly from the CPU, they feature specific instruction sets. Whereas instruction sets are quite standard on CPUs for desktop computers, there exist a variety of GPUs optimised for different purposes. Mostly due to their optimisation towards single precision operations, cards for graphics are not optimal for scientific computing applications, which are usually based on double precision computations. Cards dedicated to scientific computing are available commercially, often referred to as General-Purpose GPU (GPGPU). They can be recognised by the absence of video port. The number of single and double precision floating point operations per second (FLOPs) is usually provided by the manufacturer allowing for comparison between models.

The GPU is usually implemented as a separate card with its own memory. It is operated by the host, i.e. the computer equipped with CPUs, as illustrated in Fig.~\ref{fig-GPU}. The threads on a GPU are grouped in blocks on a grid. This structure of the threads is reflected on the memory, since only the threads in the same block have access to a shared memory.

The processing units in a GPU are called Streaming Multiprocessors (SMPs), each are composed of multiple processing units featuring high vectorisation capabilities, such that 32 or 64 threads can be executed simultaneously. Following NVIDIA's nomenclature we'll refer to the set of threads executed simultaneously by a single unit of the SMP as warps. The warps are executed independently by the computing units of the SMPs. There is a strong analogy between the execution of threads on a CPU and of warps (i.e. sets of threads) on a GPU, except that communication through a shared memory (synchronisation, locks, etc.) is possible only for threads within the same block. Only recent GPU architectures allow for block synchronisation~\cite{blockSync}.

Analogously to hyperthreading, the SMPs feature a large amount of registers allowing for the storage of multiple warps (and multiple blocks) and thus fast switching between the execution of different warps during stalls.

This architecture has some implications for the developer:
\begin{itemize}
 \item Minimum data transfer between the GPU and the host memories is advisable.
 \item The GPU performs optimally for independent identical tasks to separate datasets, at least within thread blocks. In other words, the divergence of the execution of the threads (e.g. with \textit{if} statements) within thread blocks and the communication between threads in different blocks should be minimised.
 \item Optimally the number of threads in each block should be a multiple of the warp size.
 \item Blocks featuring as many threads as possible are optimal, unless the corresponding number of blocks drops below the number of SMPs, then it is advisable to adjust the block size such that all SMPs can be active simultaneously.
\end{itemize}

There exists multiple languages to write GPU compatible codes. For example OpenCL is widely used to develop portable codes. For NVIDIA cards the corresponding proprietary language CUDA is most popular. The main difference with programming languages for CPUs is the introduction of another device with it own memory and its own source code: the kernel. The allocation of the space on the GPU memory as well as the transfer of the data and the execution of the kernel are illustrated with an example in Sec.~\ref{sec-cuda}. We note in particular the need to specify the number of blocks in the grid as well as the number of threads in each of the block.

As for CPUs, the clock speed and the memory size are important characteristics of the GPU to consider. Whereas the first is usually lower than those of CPUs, the GPU is meant to handle large datasets and thus feature significantly larger memories.
\section{The computer cluster}  \label{sec-cluster}
Computer clusters are commonly used for problem sizes beyond the capabilities of single computer. They combine the computing power of a set of individual computers, usually called nodes in this frame, by implementing a fast communication network between them thus allowing for a distribution of the tasks. Nowadays the computer clusters are usually based on Gbit/s second data transfer. These fast networks links remain $\approx$100 slower than CPU communication with its caches, such a technology is therefore appropriate only when single nodes are not sufficient. Clearly, large data transfer between nodes has to be minimised when using such a technology.

Today the most performing computer cluster is composed of 4806 individual nodes each featuring 2~CPUs and 6 GPUs to reach $\approx$200 PFLOPs (theoretical peak)~\cite{top500,summit}. However highly performing clusters are not the most common, this concept is also widely used to implement smaller machine due to it cost effectiveness. Indeed, a set of cheap desktop computers with a network switch are sufficient to build an efficient computer cluster.

Similarly to multithreading, the computing load is shared between individual processes. Unlike threads, these processes do not have access to a shared memory, since they could be executing on different nodes. From the developers' point of view, the communication between processes is handled via a Message Passing Interface (MPI)~\cite{mpi}, independently of the location of the execution of the processes, i.e. whether they are on the same node or not. The MPI consists of a set of instructions that allows communication between processes, such as transfer of data or synchronisation. An example is shown in Sec.~\ref{sec-MPI}. This example is limited to the sole usage of MPI, however multithreading or GPU acceleration can still be implemented within each node.
\subsection{Batch processing} \label{sec-batch}
Large computing resources such as computer clusters or computing grids (Sec.~\ref{sec-grid}) are usually shared between several users. In addition, the execution of heavy tasks on such machines are expected to last long, possibly up to days or even month. To deal with those constraints, it is convenient to implement an automated management of the resources. This is usually achieved using the concept of batch processing: a given task such as the execution of a program with a given input file, is described to the batch system, along with the resources required to perform the task, such as the time, the number of CPUs, the number of GPUs or even the amount of memory. Based on these information, the batch system will allocate the proper resources when available and execute the task. As opposed to the traditional usage of a personal computer, the user only submits job descriptions to the batch system, the time of the execution is no longer under his control. The allocation of the resources can then be subject to arbitrarily complex scheduling schemes, e.g. featuring equal sharing between users or favouring some according to a given payment scheme. An example of batch job description is shown in Sec.~\ref{sec-MPI}.
\section{Volunteer and grid computing}  \label{sec-grid}
For applications which do not require fast communication between processes, yet a large computing power is needed to perform a set of fully independent tasks, the usage of idling computers through a web interface becomes interesting. Most volunteer computing projects are based on the Berkeley Open Infrastructure for Network Computing (BOINC)~\cite{boinc}. Its role is to dispatch individual jobs through the web on the heterogeneous hardware made available by volunteers who subscribed to it, and retrieve the results of their execution. The main challenges linked to this type of computing is the portability of the codes which have to run on several platforms, as well as the attraction of large amounts of volunteers, which limits its usage to popular scientific projects. Due to these constraints, GPUs are often preferred for such applications.

The concept of volunteer computing became popular with the project SETI@home (Search for Extraterrestrial Intelligence)~\cite{SETI}. Today it is also used in the accelerator world, for example with the LHC@home project~\cite{lhcathome}, allowing for large scale single particle tracking studies at relatively low cost.

More generally, heavy tasks can be distributed over dedicated infrastructures, such as large computer clusters located at various laboratories or universities contributing to a given project, rather than only idling personal computers. For example, such a large scale computing grid was implemented in order to store and treat the data generated by the LHC~\cite{CERNComputingGrid}.

The interaction of the user with a computing grid is similar to the computer cluster described in Sec.~\ref{sec-batch}.

\section{Summary}
Due to the saturation of the CPU clock speed, high performance computing nowadays is achieved through parallelisation of the tasks on several computing units. We may categorise the parallelisation techniques in two main groups: SIMD and MIMD.

The most common technology implementing SIMD is multithreading (and similarly hyperthreading). From a hardware point of view, this corresponds to the combination of the computing power of multiple CPUs acting on a common memory. From a developer's point of view, this requires the writing of multithreaded codes, the usage of libraries with multithreading capabilities or the usage of compiler options to automatically parallelise loops using multiple threads. The latter comes essentially without cost on the code development side and allows for significant speed up even on common desktops. Additionally, commercial computers are available with few tens of CPUs, thus allowing for the treatment of problem sizes about an order of magnitude larger with a single machine. This parallelism is well suited for most of the algorithms used in accelerator physics. On the other hand, applications limited by memory transfer, encountered for example in the field of data analysis, may not profit from this parallelism.  GPUs, featuring a large memory bandwidth, are usually more suited then.

MIMD is implemented in most CPUs, through vectorised instructions, and in GPUs through warps. The vectorisation on commercially available CPUs can reach up to 8 double precision operations performed in parallel and can be used without major modification to a regular serial code. SMPs in commercially available GPUs allow for up to 64 operations in parallel. From the developer's point of view, the usage of GPUs usually requires an additional effort as it is based on the usage of specific languages. Nevertheless, GPU accelerated libraries as well as compiler options to enable GPU acceleration of regular codes also exist. This type of parallelism is particularly suited for independent tasks. The GPU architecture is optimal for the treatment of large datasets. In accelerators they are used for data analysis (optics corrections, processing of beam instrumentation data, machine learning), solution of differential equations (time domain electromagnetic / mechanical / thermal simulations, beam evolution), linear algebra (matrix manipulation, frequency domain electromagnetic / mechanical simulations, beam stability), Monte-Carlo (particle-matter interaction, detector design, vacuum) or beam physics via independent multiparticle tracking. The architecture of the GPU is however less suited for strongly dependent data or divergent execution of the threads, such as those encountered in Particle-in-Cell or collective effects simulations. Nevertheless, these applications require the handling of large amounts of data that is less efficient on CPUs, thus they may also profit from GPU acceleration, yet not at the level of more independent tasks. Tasks featuring small datasets are usually not well suited for GPUs, such as single particle tracking simulation, optics computations and more generally heterogeneous tasks such as those encountered in accelerator or equipment control applications. 

For tasks that exceeds the capabilities of a single machine, the computer cluster is the most common tool, combining of the resources of several machine using a MPI to communicate through a fast network connecting the nodes. For even larger scale computing projects, computing grids are used to combine the power of several computer clusters.

\bibliographystyle{unsrt}
\bibliography{CAS_ComputingTechniques}

\pagebreak
\appendix
\section{Source codes} \label{sec-sources}
The sources codes implementing the same algorithm but parallelised with five different technologies are shown in the following. The codes are written in \textit{C} aiming at being explicit more than performant. We note that these technologies are also available in many other languages with slightly different syntaxes. Also, many libraries implement one or more possibilities for parallelism that can be enabled if setup properly for a given hardware. Consequently it is clear that this set of examples is far from providing a complete overview of parallelised computing techniques, but rather aims at easing the readers' first steps into this field.

\begin{figure}[b!]
 \begin{center}
  \includegraphics[width=0.6\linewidth]{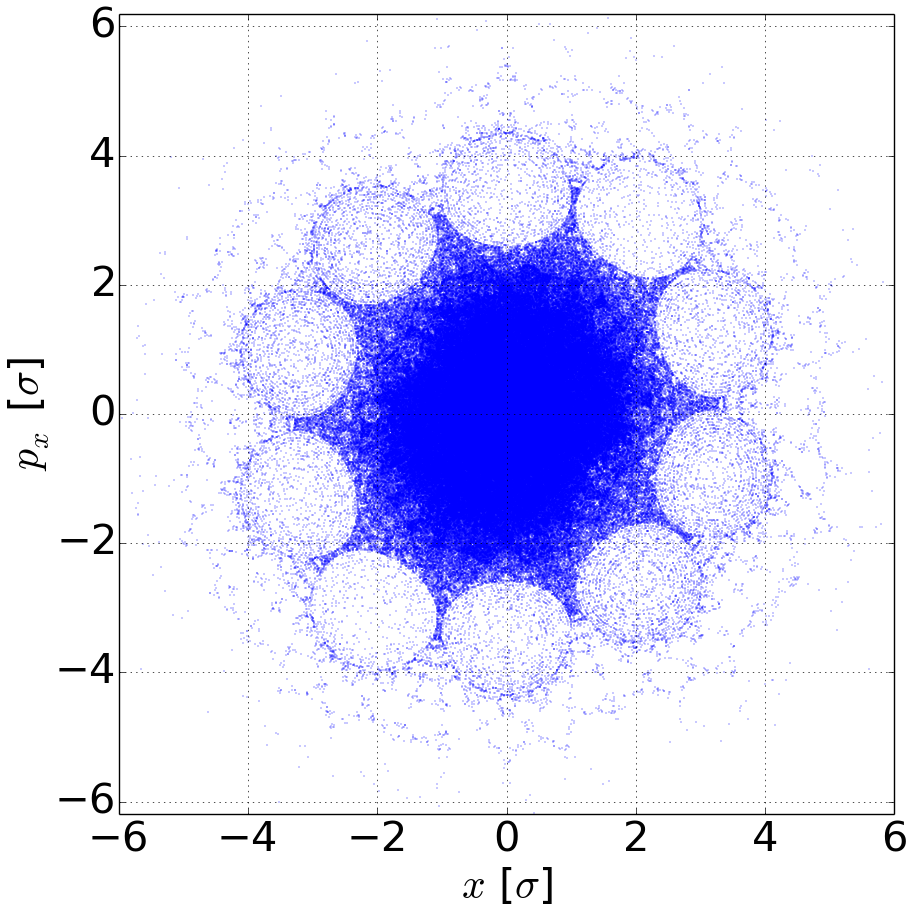}
 \end{center}
\caption{Position in phase space of $5\cdot10^{5}$ particles with different initial conditions after $10^4$ turns in a circular collider with a strong head-on beam-beam interaction. The islands generated by the 10$^{\text{th}}$ order resonance are visible, along with more complex substructures.}
\label{fig-phaseSpace}
\end{figure}

The example is taken from the study of non-linear dynamics, in particular the impact of the electromagnetic interaction between the beams in a circular collider. The result of the execution is shown in Fig.~\ref{fig-phaseSpace}.

\pagebreak
\subsection{Serial / OpenMP} \label{sec-openMP}
To enable multithreading with OpenMP using gcc, the flag \textit{-fopenmp} is required at compilation.\\

\begin{lstlisting}[language=c,numbers=left,basicstyle=\fontsize{8}{2}\ttfamily,commentstyle=\color{blue}]
#include<math.h>
#include<cstdio>
#include<stdlib.h>

int main(int argc, char *argv[]){
    // Allocation of the memory for the particles coordinates
    const int nPart = 500000;
    double x[nPart];
    double px[nPart];
    // Initiallisation of the particles coordinates with Gaussian distributions
    double radius;
    double angle;
    for(int i = 0;i<nPart;++i){
        radius = (double)rand() / RAND_MAX;
        while(radius==0){
            radius = rand();
        }
        radius = sqrt(-2.0*log(radius));
        angle = (double)2.0*M_PI*rand() / RAND_MAX;
        x[i] = radius*sin(angle);
        px[i] = radius*cos(angle);
    }
    // Tracking variables
    int nTurn = 10000;
    double sinPhix = sin(2.0*M_PI*0.31);
    double cosPhix = cos(2.0*M_PI*0.31);
    double scale = 1.0;
    double thres = 1E-10;
    double oldX = 1.0;
    double oldPx = 1.0;
    // Tracking
    for(int turn = 0;turn<nTurn;++turn){
        // Multithreading with the large arrays shared and the tracking variables
        // copied for each thread (firstprivate). The size of the threads
        // is adaptive starting at 1000 particles per thread (schedule).
        #pragma omp parallel for default(none) shared(x,px) \
                                 firstprivate(oldX,oldPx,cosPhix,sinPhix,thres,scale) \
                                 schedule(guided,1000)
        for(int i = 0;i<nPart;++i){
            // Phase space rotation
            oldX = x[i];
            oldPx = px[i];
            x[i] = cosPhix*oldX + sinPhix*oldPx;
            px[i] = -sinPhix*oldX + cosPhix*oldPx;
            // Beam-beam kick
            if(abs(x[i]) > thres){
                px[i] += scale*(1.0-exp(-0.5*x[i]*x[i]))/x[i];
            }
        }
    }
    // Output the particles coordinates to a file
    FILE* file = fopen("phaseSpace_OMP.csv","w");
    for(int i=0;i<nPart;++i){
        fprintf(file,"%E,%E\n",x[i],px[i]);
    }
    fclose(file);
}
\end{lstlisting}
\pagebreak
\subsection{POSIX threads} \label{sec-pthreads}
To enable multithreading with POSIX thread using gcc, the flag \textit{-pthread} is required at compilation.\\

\begin{lstlisting}[language=c,numbers=left,basicstyle=\fontsize{8}{2}\ttfamily,commentstyle=\color{blue}]
#include<math.h>
#include<cstdio>
#include<stdlib.h>
#include <pthread.h>

// Allocation of the memory for the particles coordinates
// as global variables
const int nPart = 500000;
double x[nPart];
double px[nPart];
// Tracking variables defined as global
int nTurn = 10000;
double sinPhix = sin(2.0*M_PI*0.31);
double cosPhix = cos(2.0*M_PI*0.31);
double scale = 1.0;
double thres = 1E-10;
// Definition of the code to be executed by each thread
void* track(void* indices_ptr){
    int* indices = (int*) indices_ptr;
    double oldX = 0.0;
    double oldPx = 0.0;
    // Tracking
    for(int turn = 0;turn<nTurn;++turn){
        for(int i = indices[0];i<indices[1];++i){
            // Phase space rotation
            oldX = x[i];
            oldPx = px[i];
            x[i] = cosPhix*oldX + sinPhix*oldPx;
            px[i] = -sinPhix*oldX + cosPhix*oldPx;
            // Beam-beam kick
            if(abs(x[i]) > thres){
                px[i] += scale*(1.0-exp(-0.5*x[i]*x[i]))/x[i];
            }
        }
    }
};
int main(int argc, char *argv[]){
    // Initiallisation of the particles coordinates with Gaussian distributions
    double radius;
    double angle;
    for(int i = 0;i<nPart;++i){
        radius = (double)rand() / RAND_MAX;
        while(radius==0){
            radius = rand();
        }
        radius = sqrt(-2.0*log(radius));
        angle = (double)2.0*M_PI*rand() / RAND_MAX;
        x[i] = radius*sin(angle);
        px[i] = radius*cos(angle);
    }
    // Definition of the chunks of particles for each thread
    int indices0[2] = {0,100000};
    int indices1[2] = {400000,500000};
    int indices2[2] = {100000,200000};
    int indices3[2] = {200000,300000};
    int indices4[2] = {300000,400000};
    // Creation of 4 additional threads
    pthread_t thread1;
    pthread_t thread2;
    pthread_t thread3;
    pthread_t thread4;
    pthread_create(&thread1, NULL, track, indices1);
    pthread_create(&thread2, NULL, track, indices2);
    pthread_create(&thread3, NULL, track, indices3);
    pthread_create(&thread4, NULL, track, indices4);
    // Tracking on the current thread, executing in parallel 
    // to the other 4.
    track(indices0);
    // Synchronisation of the threads with the current one
    pthread_join(thread1, NULL);
    pthread_join(thread2, NULL);
    pthread_join(thread3, NULL);
    pthread_join(thread4, NULL);
    // Output the particles coordinates to a file once all threads are synchronised, 
    // i.e. have finished their execution
    FILE* file = fopen("phaseSpace_phtreads.csv","w");
    for(int i=0;i<nPart;++i){
        fprintf(file,"%E,%E\n",x[i],px[i]);
    }
    fclose(file);
}
\end{lstlisting}
\pagebreak
\subsection{CUDA} \label{sec-cuda}
CUDA codes require compilation with a dedicated compiler (nvcc).\\

\begin{lstlisting}[language=c,numbers=left,basicstyle=\fontsize{8}{2}\ttfamily,commentstyle=\color{blue}]
#include<math.h>
#include<cstdio>
#include<stdlib.h>

// Kernel definition (single turn tracking)
__global__
void track(double* x, double* px,int nPart, double sinPhix,double cosPhix, \
           double scale,double thres)
{
    // Mapping of the block-thread indices to 1D array indices
    int i = blockIdx.x*blockDim.x + threadIdx.x;
    // Phase space rotation
    double oldX = 0.0;
    double oldPx = 0.0;
    oldX = x[i];
    oldPx = px[i];
    x[i] = cosPhix*oldX + sinPhix*oldPx;
    px[i] = -sinPhix*oldX + cosPhix*oldPx;
    // Beam-beam kick
    if(abs(x[i]) > thres){
        px[i] += scale*(1.0-exp(-0.5*x[i]*x[i]))/x[i];
    }
}

int main(int argc, char *argv[]){
    // Allocation of the memory for the particles coordinates on the host memory
    const int nPart = 500000;
    double x[nPart];
    double px[nPart];
    // Initiallisation of the particles coordinates with Gaussian distributions
    // on the host memory
    double radius;
    double angle;
    for(int i = 0;i<nPart;++i){
        radius = (double)rand() / RAND_MAX;
        while(radius==0){
            radius = rand();
        }
        radius = sqrt(-2.0*log(radius));
        angle = (double)2.0*M_PI*rand() / RAND_MAX;
        x[i] = radius*sin(angle);
        px[i] = radius*cos(angle);
    }
    // Allocation of the memory for the particles coordinates on the GPU memory
    double* x_cuda;
    double* px_cuda;
    cudaMalloc(&x_cuda, nPart*sizeof(double)); 
    cudaMalloc(&px_cuda, nPart*sizeof(double));
    // Copy of the particles coordinates to the GPU memory
    cudaMemcpy(x_cuda, x, nPart*sizeof(double), cudaMemcpyHostToDevice);
    cudaMemcpy(px_cuda, px, nPart*sizeof(double), cudaMemcpyHostToDevice);
    // Definition of the thread structure on the GPU
    int threadsperblock = 512;
    int blockspergrid = ceil(nPart / threadsperblock);
    // Tracking variables
    int nTurn = 10000;
    double sinPhix = sin(2.0*M_PI*0.31);
    double cosPhix = cos(2.0*M_PI*0.31);
    double scale = 1E-4;
    double thres = 1E-10;
    // Tracking
    for(int turn = 0;turn<nTurn;++turn){
        track<<<blockspergrid,threadsperblock>>>(x_cuda,px_cuda,nPart,sinPhix,cosPhix, \
                                                 scale,thres);
    }
    // Copy the particles coordinates back to the host memory
    cudaMemcpy(x, x_cuda, nPart*sizeof(double), cudaMemcpyDeviceToHost);
    cudaMemcpy(px, px_cuda, nPart*sizeof(double), cudaMemcpyDeviceToHost);
    // deallocate the GPU memory
    cudaFree(x_cuda);
    cudaFree(px_cuda);
    // Output the particles coordinates to a file
    FILE* file = fopen("phaseSpace_cuda.csv","w");
    for(int i=0;i<nPart;++i){
        fprintf(file,"%E,%E\n",x[i],px[i]);
    }
    fclose(file);
}
\end{lstlisting}
\pagebreak
\subsection{OpenACC} \label{sec-openACC}
To enable GPU acceleration using \textit{gcc} and \textit{OpenACC}~\cite{openACC}, the flag \textit{-fopenacc} is required at compilation, along with the path to a CUDA compiler to which the parallelised part is offloaded, e.g. \textit{-foffload=nvptx-none}\\

\begin{lstlisting}[language=c,numbers=left,basicstyle=\fontsize{8}{2}\ttfamily,commentstyle=\color{blue}]
#include<math.h>
#include<cstdio>
#include<stdlib.h>

int main(int argc, char *argv[]){
    // Allocation of the memory for the particles coordinates
    const int nPart = 500000;
    double x[nPart];
    double px[nPart];
    // Initiallisation of the particles coordinates with Gaussian distributions
    double radius;
    double angle;
    for(int i = 0;i<nPart;++i){
        radius = (double)rand() / RAND_MAX;
        while(radius==0){
            radius = rand();
        }
        radius = sqrt(-2.0*log(radius));
        angle = (double)2.0*M_PI*rand() / RAND_MAX;
        x[i] = radius*sin(angle);
        px[i] = radius*cos(angle);
    }
    // Tracking variables
    int nTurn = 10000;
    double sinPhix = sin(2.0*M_PI*0.31);
    double cosPhix = cos(2.0*M_PI*0.31);
    double scale = 1.0;
    double thres = 1E-10;
    double oldX = 1.0;
    double oldPx = 1.0;
    // Tracking, to be treated as GPU Kernel
    #pragma acc parallel
    for(int turn = 0;turn<nTurn;++turn){
        #pragma acc loop
        for(int i = 0;i<nPart;++i){
            // Phase space rotation
            oldX = x[i];
            oldPx = px[i];
            x[i] = cosPhix*oldX + sinPhix*oldPx;
            px[i] = -sinPhix*oldX + cosPhix*oldPx;
            // Beam-beam kick
            if(abs(x[i]) > thres){
                px[i] += scale*(1.0-exp(-0.5*x[i]*x[i]))/x[i];
            }
        }
    }
    // Output the particles coordinates to a file
    FILE* file = fopen("phaseSpace_OACC.csv","w");
    for(int i=0;i<nPart;++i){
        fprintf(file,"%E,%E\n",x[i],px[i]);
    }
    fclose(file);
}
\end{lstlisting}

\pagebreak
\subsection{MPI}  \label{sec-MPI}
The implementations of the MPI feature both a compiler, e.g. \textit{mpicc}, and a run command such as \textit{mpirun} or \textit{mpiexec}. The latter requires at least the number of processes to be specified when running the code, e.g. using the flag \textit{-np}. This command will launch the corresponding numbers of identical copies of the executable, but featuring a different rank.\\

\begin{lstlisting}[language=c,numbers=left,basicstyle=\fontsize{8}{2}\ttfamily,commentstyle=\color{blue}]
#include<math.h>
#include<cstdio>
#include<stdlib.h>
#include "mpi.h"

int main(int argc, char *argv[]){
    // Definition of tags for MPI calls (optional)
    int TAGREADY = 0;
    int TAGX = 1;
    int TAGPX = 2;
    // Allocation of the memory for the particles coordinates
    // (for each process)
    const int nPart = 500000;
    double x[nPart];
    double px[nPart];
    // MPI initialisation, retrival of the process rank and the total
    // number of processes
    MPI_Init(&argc, &argv);
    int myRank;
    int activeProcs;
    MPI_Comm_rank(MPI_COMM_WORLD, &myRank);
    MPI_Comm_size(MPI_COMM_WORLD, &activeProcs);
    int chunkSize = ceil(nPart/activeProcs);
    if(myRank == 0){
        // Initiallisation of the particles coordinates with Gaussian distributions
        // performed by one process only
        double radius;
        double angle;
        for(int i = 0;i<nPart;++i){
            radius = (double)rand() / RAND_MAX;
            while(radius==0){
                radius = rand();
            }
            radius = sqrt(-2.0*log(radius));
            angle = (double)2.0*M_PI*rand() / RAND_MAX;
            x[i] = radius*sin(angle);
            px[i] = radius*cos(angle);
        }
        // Send a subset of the initial coordinates to other processes
        for(int iProc=1;iProc<activeProcs;++iProc){
            MPI_Send(&x[iProc*chunkSize], chunkSize, MPI_DOUBLE, iProc,TAGX,   \
                     MPI_COMM_WORLD);
            MPI_Send(&px[iProc*chunkSize], chunkSize, MPI_DOUBLE, iProc,TAGPX, \
                     MPI_COMM_WORLD);
        }
    } else{
        // Processes with other ranks wait until they receive their subset of
        // initial particles coordinate 
        MPI_Status status;
        MPI_Recv(x,chunkSize, MPI_DOUBLE, 0,TAGX, MPI_COMM_WORLD, &status);
        MPI_Recv(px,chunkSize, MPI_DOUBLE, 0,TAGPX, MPI_COMM_WORLD, &status);
    }
    // Tracking variables (executed by all processes)
    int nTurn = 10000;
    double sinPhix = sin(2.0*M_PI*0.31);
    double cosPhix = cos(2.0*M_PI*0.31);
    double scale = 1.0;
    double thres = 1E-10;
    double oldX = 0.0;
    double oldPx = 0.0;
    // Tracking (executed by all processes)
    for(int turn = 0;turn<nTurn;++turn){
        for(int i = 0;i<chunkSize;++i){
            oldX = x[i];
            oldPx = px[i];
            x[i] = cosPhix*oldX + sinPhix*oldPx;
            px[i] = -sinPhix*oldX + cosPhix*oldPx;
            if(abs(x[i]) > thres){
                px[i] += scale*(1.0-exp(-0.5*x[i]*x[i]))/x[i];
            }
        }
    }
    if(myRank==0){
        // One process collects the particles coordinates computed by other processes
        int nRecv = 1;
        while(nRecv<activeProcs){
            MPI_Status status;
            MPI_Recv(NULL, 0, MPI_BYTE, MPI_ANY_SOURCE,TAGREADY, MPI_COMM_WORLD,&status);
            int iProc = status.MPI_SOURCE;
            MPI_Recv(&x[iProc*chunkSize], chunkSize, MPI_DOUBLE, status.MPI_SOURCE,TAGX, \
                     MPI_COMM_WORLD,&status);
            MPI_Recv(&px[iProc*chunkSize], chunkSize, MPI_DOUBLE, status.MPI_SOURCE,TAGPX, \
                     MPI_COMM_WORLD,&status);
            ++nRecv;
        }
    } else{
        // All other processes send their data after tracking
        MPI_Send(NULL, 0, MPI_BYTE, 0,TAGREADY, MPI_COMM_WORLD);
        MPI_Send(x,chunkSize, MPI_DOUBLE, 0,TAGX, MPI_COMM_WORLD);
        MPI_Send(px,chunkSize, MPI_DOUBLE, 0,TAGPX, MPI_COMM_WORLD);
    }
    MPI_Finalize();
    // Only the process that has collected all the data outputs the particles 
    // coordinates to a file
    if(myRank==0){
        FILE* file = fopen("phaseSpace_MPI.csv","w");
        for(int i=0;i<nPart;++i){
            fprintf(file,"%E,%E\n",x[i],px[i]);
        }
        fclose(file);
    }
}
\end{lstlisting}
\vspace{2mm}
When using batch processing, the syntax for the submission of the job depends on the technology. Below an example of a script for SLURM~\cite{slurm} illustrates the request for resources (here 20 nodes are requested, each running 40 processes for 10 minutes). The \textit{mpirun}/\textit{mpiexec} and the corresponding flags which are used in particular for the allocation of the resources are replace by the command \textit{srun}, as the allocation is now under the control of the batch system. The submission to the scheduler is done via a given command, here \textit{sbatch jobFileName}.\\

\begin{lstlisting}[language=bash,numbers=left,basicstyle=\fontsize{8}{2}\ttfamily,commentstyle=\color{blue}]
#!/bin/bash
#SBATCH --nodes 20
#SBATCH --tasks-per-node 40
#SBATCH --cpus-per-task 1
#SBATCH --time 00:10:00
#SBATCH --workdir /pathTo/phaseSpace

srun phaseSpace_MPI
\end{lstlisting}

\end{document}